\documentclass[manuscript]{acmart}

\usepackage{tabularx}
\usepackage{multirow}

\graphicspath{{images/}}
\usepackage[caption=false]{subfig}

\AtBeginDocument{%
  \providecommand\BibTeX{{%
    \normalfont B\kern-5em{\scshape i\kern-25em b}\kern-8em\TeX}}}

\setcopyright{rightsretained}

\begin{document}

\title{Towards Sustainable Research Data Management in Human-Computer Interaction}

\author{David Goedicke}
\email{dg536@cornell.edu}
\orcid{0000-0002-4837-893X}
\affiliation{%
  \institution{Cornell University}
  \streetaddress{2 W Loop Road}
  \city{New York}
  \country{U.S.}}

\author{Mark Colley}
\email{mark.colley@uni-ulm.de}
\orcid{0000-0001-5207-5029}
\affiliation{%
  \institution{Institute of Media Informatics, Ulm University}
  \city{Ulm}
  \country{Germany}
}
\affiliation{%
  \institution{Cornell Tech}
  \streetaddress{2 W Loop Road}
  \city{New York}
  \country{U.S.}
}

\author{Sebastian S. Feger}
\email{sebastian.feger@ifi.lmu.de}
\orcid{0000-0002-0287-0945}
\affiliation{%
  \institution{LMU Munich}
  \streetaddress{Frauenlobstr. 7a}
  \city{Munich}
  \country{Germany}}

\author{Michael Goedicke}
\email{michael.goedicke@paluno.uni-due.de}
\orcid{}
\affiliation{%
  \institution{University of Duisburg-Essen}
  \streetaddress{}
  \city{Essen}
  \country{Germany}}

\author{Bastian Pfleging}
\email{bastian.pfleging@informatik.tu-freiberg.de}
\orcid{0000-0003-0505-9338}
\affiliation{%
  \institution{TU Bergakademie Freiberg}
  \streetaddress{}
  \city{Freiberg}
  \country{Germany}}

\author{Wendy Ju}
\email{wendyju@cornell.edu}
\orcid{0000-0002-3119-611X}
\affiliation{%
  \institution{Cornell University}
  \streetaddress{2 W Loop Road}
  \city{New York}
  \country{U.S.}}


\begin{abstract}
We discuss important aspects of HCI research regarding Research Data Management (RDM) to achieve better publication processes and higher reuse of HCI research results. Various context elements of RDM for HCI are discussed, including examples of existing and emerging infrastructures for RDM. We briefly discuss existing approaches and come up with additional aspects which need to be addressed. This is to apply the so-called FAIR principle fully, which -- besides being findable and accessible -- also includes interoperability and reusability. We also discuss briefly the kind of research data types that play a role here and propose to build on existing work and involve the HCI scientific community to improve current practices. 

\end{abstract}

\ccsdesc[300]{General and reference}


\keywords{Research Data Management in HCI, Open Science, HCI community, Sustainability of HCI research}

\maketitle

\section{Introduction}
Reproducibility is a cornerstone of science that drives validation, trust, and reusability. These scientific characteristics are also of great interest to the Human-Computer Interaction (HCI) community, as evidenced by past contributions and initiatives~\cite{10.1145/1979742.1979491, 10.1145/3170427.3185377}. HCI researchers even described and stressed the special role of scientists and practitioners in this community to design systems that make much-needed systematic research data management (RDM) together with reuse, possible and accessible while providing strong incentives and rewards for participating community members~\cite{10.1145/3290607.3312905}. Yet, given the broad range of methods in HCI, most initiatives focus on individual research directions. One example is \textit{Transparent Statistics in Human–Computer Interaction}~\cite{transparentstatistics}, which focuses primarily on quantitative research data. In this paper, we aim to engage in a discussion on how to use and re-purpose existing and emerging RDM infrastructure(s) to match a large spectrum of HCI methods under common systems and interfaces. To this end, we particularly relate to the following timely topics in this paper: 1) profiting from existing national and international initiatives within and beyond computer science (CS); 2) the use of common RDM systems in HCI research; 3) the adoption of science badges that reward reproducibility efforts; and 4) the introduction of automated FAIR (Findable, Accessible, Interoperable, and Reusable) scoring.

The general state of affairs regarding RDM and the proclaimed FAIR~\cite{gofair}- principles (Find, Access, Interoperate, Reuse) is that there are several initiatives also at the general level, which range from recommendations on procedures and plans/efforts to build entire infrastructures for specific scientific disciplines. In CS, data is often already digitized, so it is rather easy to make research data (and related artifacts and tools) \textit{accessible} by putting it into some repository or onto a web page. With sufficient informal metadata describing it, this might suffice for the \textit{Find}-property for the research data is also available since general web search engines can index it. 

While finding and accessing HCI research seems manageable, \textit{interoperation} and \textit{reuse} are still elusive. 
This is because these aspects need more work regarding the research data. They imply that more effort is spent on metadata specification and long-term archival. 
For interoperation, common metadata standards need to be developed and agreed upon. For reuse to be possible, we need longer-lasting research infrastructure
-- currently, a period of about 10 years is considered here~\cite{dfg2}. 

Thus, these aspects need efforts across the scientific communities, also for establishing and running infrastructure. Overarching organizations like scientific societies or basic research funding agencies are needed to provide organizational and financial resources to put these forward. This has been recognized in many countries, and related efforts have been started across the globe (e.g., in the Netherlands\footnote{\url{https://www.surf.nl/en/national-coordination-point-research-data-management}, last access: 2023-01-18} and Germany\footnote{\url{https://www.nfdi.de/}, last access: 2023-01-18}).

From our perspective, attempting to tackle these challenges means that the communities have to support these efforts by defining various standards (metadata types, research data types, quality standards, etc.) to serve the community and not the peculiar/specific interests of other market players. The question is what this means for the HCI community and how a way forward can be spelled out.

To develop first ideas to pick up the existing work and extend these to push forward in the direction of those goals, we discuss typical research data types of HCI research. We then review various known efforts of frameworks to implement such ideas in the form of actual processes and infrastructures. We also discuss the various efforts like the badge system, which is a start to include and encourage the storage of research data and artifacts in publications. Currently, the use of these badges is slowly encouraged/enforced by the publication processes. Badges support a part of the aspects of the FAIR principles. It will be interesting to assess to which degree this supports these principles. Ways to evaluate this degree need to be community-specific, and we discuss this also briefly before we embark on a discussion of how to propose ways to go forward and increase the support by and for the researchers to come up with better RDM, and more publications, citations that foster reuse as well. Additionally, we provide an analysis of HCI publications regarding their use of methods for open-sourcing, such as preregistration or employing GitHub.

\subsection{Research Data in HCI} 
\citet{ArtifactTransp2020} highlighted many challenges about the various data types, artifacts, and policies typically used in HCI. We elaborate on these as HCI has specific requirements and issues for RDM over CS.

\subsubsection{Human Participant Data} 
First and foremost, HCI papers and research often feature human participants and often feature data that might be personally identifying or which might cause personal jeopardy if revealed in the wrong context. These include photos and videos of the participants of their behaviors (e.g.,~\cite{10.1145/2809730.2809752, 10.1145/3491102.3517571}), their written or verbal responses (e.g.,~\cite{colley2022requirements}) to questionnaires, logs from environmental or wearable sensors (e.g.,~\cite{10.1145/3534587}), physiological data from on-body or remote sensors, and physical location traces. 

Naturalistic data capture is often desirable to help understand people's behavior, which might be relevant to the design of user-centered or context-sensitive inventions or provide a pre-condition against which post-intervention outcomes might be compared. However, naturalistic data capture also often captures unintended phenomena and activities; these can be both convenient for secondary analyses and reveal information that participants or researchers did not intend to share.

Data captured from controlled experiments or other studies with interventions may show behaviors or reactions from people that, taken out of context, could affect the participants. For example, a professional driver driving poorly in a simulation experiment might jeopardize their professional livelihood. If mishandled, physiological data obtained from a controlled study could potentially negatively impact the insurability of a patient or reveal biometric features that need to be protected.

\subsubsection{Artifacts} 
 A significant portion of CS work directly tests or tangentially uses artifacts as part of their paper's contribution. This could be an implementation of an algorithm, a data set, or data analysis code. HCI work similarly produces software artifacts that need to become part of the archived work. However, HCI frequently involves physical artifacts, including controllers, interfaces, displays, or systems through which users interact. The design files for such physical artifacts also need to be stored, along with associated data or files such as circuit schematics, printed circuit board (PCB) designs, Bill Of Materials (BOM), mechanical part schematics, micro-controller code, and so on--anything required to reproduce the work fully. In some cases, even specialized hardware designs might be emulated by the technology available via QEMU~\cite{qemu}.

\subsubsection{Contextual information}
HCI work is often highly contextual; hence, an important aspect of capturing HCI is recording information about the temporal, physical, geographical, or cultural context in which work occurs. This could include ethnographic-style photos or videos of the location where HCI work is studied or intended to be deployed.

\subsection{Research Data Management in HCI} 
Long-term storage and availability of research data is part of what RDM is supposed to provide for the scientific community. The ability to go back, reanalyze, and verify data and artifacts are important next steps to advance reproducibility within the community. This feature of RDM, however, conflicts directly with some types of HCI research data. In particular, this covers any data that can be used to identify a person. This is referred to as Personal Identifiable Information (PII). Data that contains PII is often subject to data-privacy laws that vary significantly across the globe. The most restrictive privacy protection regulations, GDPR~\cite{eu2018gdpr} in Europe, make long-term RDM difficult for HCI research. 

Laws like GDPR lead to uncertainty and cautious researchers~\cite{ArtifactTransp2020} that delete data early or choose not to collect certain data types at all. This behavior might technically lead to GDPR compliance, but it might be overly cautious as GDPR does accommodate research.  In particular, work that provides a public good (see Art. 6 Point 5~\cite{eu2018gdpr}) or is in the legitimate interest of the expressed research goals (see Art. 6 Point 6~\cite{eu2018gdpr}) is often allowed under GDPR. One essential feature of a GDPR-compliant RDM infrastructure will be its flexibility to allow for later deletion requests by individual participants. Therefore, the scenario is not expected to be overly unfavorable. Given both these restrictions and accommodations within the GDPR laws, HCI research can still be done with a meaningful collection and retention of data.


From larger data to long-term modification access, HCI presents a wide range of specific challenges to long-term RDM solutions. And while it is generally possible to archive data, these particular regiments are often not supported, which leads to the opportunity to develop and deploy an RDM standard for the HCI community. 

\section{Existing Methods and Opportunities for RDM}
RDM is currently being explored, primarily by local or national efforts. In this section, we highlight approaches deployed in Germany and the US as two examples but also examine gaps in these existing efforts. Many other efforts also exist (e.g., SURF~\cite{surf} in the Netherlands).

\subsection{National Science Foundation and US Efforts} 
In the US, many individual institutions have their own digital repositories 
(e.g., Cornell eCommons~\cite{eommons}, Stanford Digital Repository~\cite{stanford}, MIT's DSpace~\cite{dspace}.) Institutional repositories have been noted by information science scholars to be underutilized due to lack of awareness, concerns about whether repository use constitutes "publishing", and lack of external incentives and motivation~\cite{reid2008investigation, davis2007institutional}. These repositories are not organized systematically across institutions, which makes it difficult for what research data is stored to be indexed and accessed by search engines or meta-interfaces.

The National Science Foundation (NSF) has an initiative called Cyberinfrastructure for Sustained Scientific Innovation~\cite{NSFCyber}, which includes funding awards for service elements, framework implementations, and transition to sustainability resulting in a community framework providing CI services to support scientific innovation. This would be the appropriate program under which to develop RDM tools for HCI. To date, there is a funded project for ``Cyberinfrastructure for Human-Robot Interaction Research'' led by Jeffrey Hansen~\cite{NSFROBOTS}, but a similar project for HCI research is missing. 

\subsection{Nationale Forschungsdateninfrastruktur}
The Nationale Forschungssdateninfrastruktur~\cite{dfg2} (National Research Data Infrastructure / Initiative) in Germany is a 10-year funding scheme that identifies scientific communities and supports these by 90 m€ per year to build discipline-specific infrastructures for related research data repositories. This will address support for procedures, processes, and definitions but not actual storage and computing power to operate repositories. Meanwhile, the three-round proposal phase is closed, and 27 consortia have been selected. Funding is secured for the first phase of these projects for the next five years (continuation has to be proposed for the second phase soon). Two consortia will address CS and its applications. The consortium NFDIxCS~\cite{nfdixcs} with 18 partners and about 30 further supporting individuals address CS and its sub-disciplines properly, while NFDI4Datascience~\cite{NFDI4ds} will support the application of CS methods (and the related RDM) in certain application areas. 

HCI is covered in NFDIxCS and is one of the starting sub-disciplines leading the way into the general support of CS RDM. The envisioned architecture and concepts have been outlined and summarized by \citet{mci/Goedicke2022}. The specific NFDIxCS approach addresses governance, processes, and architectures to support RDM for CS per sub-discipline by using the so-called RDMC (RDM Container). This serves as a common structure for a kind of time capsule to pack all the research data (raw, purged, processed, analyzed, and presented) [publication] metadata and context in the form of execution environment additional software and libraries and support information into one closed object – very similar to the currently discussed concept of the fair data objects~\cite{fairdo}. In addition, the RDMC contains access control and workflows to support the various processes proposed by either the RDM plan or the publication processes managing the quality of the research data in question. The project will start in the spring of 2023, and first prototypes are planned for the end of 2023. 


\section{Adoption of Science Badges}

\begin{figure}[ht]
  \centering
    \subfloat[][]{
        \includegraphics[height=0.21\columnwidth]{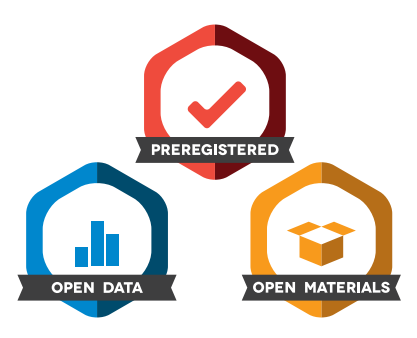}
        \label{fig:osb}}
    \subfloat[][]{
        \includegraphics[height=0.21\columnwidth]{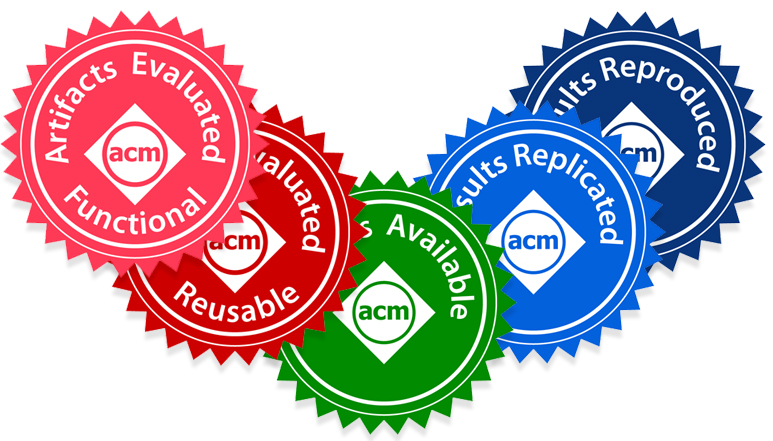}
        \label{fig:acmbadges}}
  \caption{Badges used to incentivize and reward comprehensive RDM and openness. (a) Generic Open Science badges; (b) ACM badges.}
  \label{fig:badges1}
\end{figure}

\begin{figure}[ht]
  \centering
    \subfloat[][]{
        \includegraphics[height=0.3\columnwidth]{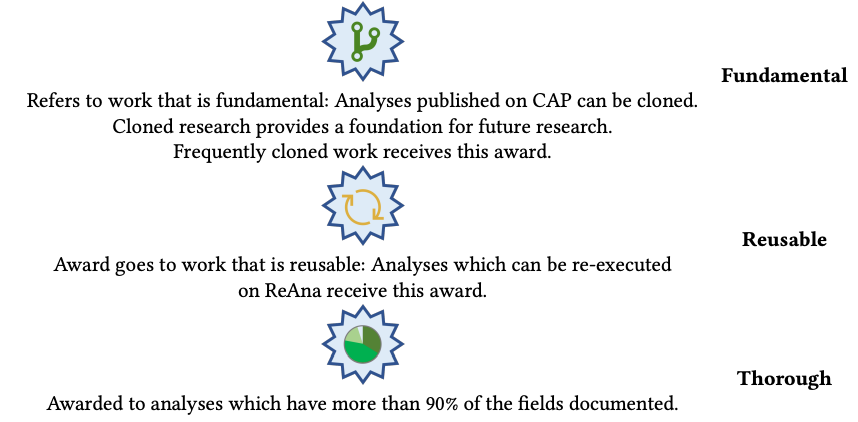}
        \label{fig:tailored}}
    \subfloat[][]{
        \includegraphics[height=0.30\columnwidth]{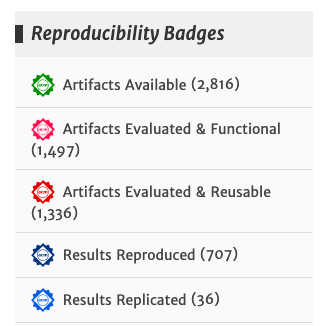}
        \label{fig:acmDLbadges}}
  \caption{(a) Tailored Science Badges proposed to match reproducible research requirements in particle physics; (b) Overview of the ACM badges on the ACM Digital Library for the search query "Open Science HCI".}
  \label{fig:badges2}
\end{figure}

Science badges incentivizing and rewarding comprehensive RDM and data openness are gaining increased attention. They can broadly be categorized across a spectrum ranging from \textit{generic} mechanisms to badges that are closely \textit{tailored} to individual scientific fields and institutions. \autoref{fig:osb} depicts the Open Science Badges (OSB) from the Center for Open Science. OSB are highly generic badges rewarding preregistration, data openness, and material openness. Their generic nature led to the badge adoption across 75 journals from a range of different scientific fields. Studying the effect of OSB, \citet{kidwell2016badges} found that the data sharing rates in a psychological science journal increased strongly after adopting the badges, as compared to similar journals not doing so. In contrast, \autoref{fig:acmbadges} depicts the ACM Reproducibility Badges that are more closely tailored to the computational research within ACM's scope while maintaining a generic nature that allows to adopt them across different venues. \citet{tailored_badges} designed even more tailored badges that match the requirements of particle physics research at CERN. Given these initiatives and indications that well-designed science badges can be effective, the HCI community might discuss whether and how to adopt them. The ACM badges could clearly be the most accessible ones. \autoref{fig:acmDLbadges} shows a search result page in the ACM Digital Library that can be filtered according to badge types. This could also increase the visibility of HCI research.

\section{Analysis of Availability in HCI}
To evaluate how many papers already include some form of the FAIR principles, we used a script-based approach to go over all papers from
W4A~\cite{w4a}, TACCESS~\cite{taccess}, CHI~\cite{chi}, UIST~\cite{uist}, AutoUI~\cite{autoui}, and MUM~\cite{mum} in the frame of 2016 -- 2019. 
We analyzed these publications via an R script using the \textit{pdfsearch}~\cite{brandon2019package}. We defined relevant terms to search the publications: "github.com", "gitlab", "osf.io", "preregistration", "GDPR", "source code", "research data management", "openly available", "will be made available", "MIT License", "cos.io", "plos.org", "https://aspredicted.org/", "re3data.org", "https://zenodo.org/". These terms were based on their own approaches and those mentioned by \citet{ArtifactTransp2020}.
The results are shown in \autoref{tab:analysis}. The values range between 0\% and 44\%.

\begin{table}[ht]
\caption{Analysis of papers. Percentage rounded.}
\small
\label{tab:analysis}
\begin{tabular}{l|lllllll}
\hline
\textbf{Year} & \textbf{ASSETS} & \textbf{AutoUI} & \textbf{CHI} & \textbf{MUM} & \textbf{TACCESS} & \textbf{UIST} & \textbf{W4A} \\ \hline
2019 & 7/41 (17\%)   & 4/34 (12\%)  & 144/703  (20\%)    & 7/38 (18\%)        & 3/11  (27\%)  & 30/95   (32\%)      & 8/18   (44\%)      \\
2018 & 8/28 (29\%)  & 3/35 (09\%)  & 123/665 (18\%)     & 8/38 (21\%)        & 5/20  (25\%)  & 31/80    (39\%)     & 1/24   (04\%)      \\
2017 & 4/28  (14\%) & 1/30  (03\%) & 79/599 (13\%)      & 3/38 (08\%)        & 6/23  (26\%)  & 15/73 (21\%)        & 6/31   (19\%)      \\
2016 & 1/24 (04\%)     & 0/37  (00\%) & 66/545 (12\%)      & 1/34 (03\%)        & 3/16  (19\%)  & 13/79  (16\%)       & 5/35   (14\%)      \\ \hline
\end{tabular}
\end{table}

\section{Use of Automated FAIR scoring}

\begin{figure}[ht]
    \centering
    \includegraphics[width=0.5\textwidth]{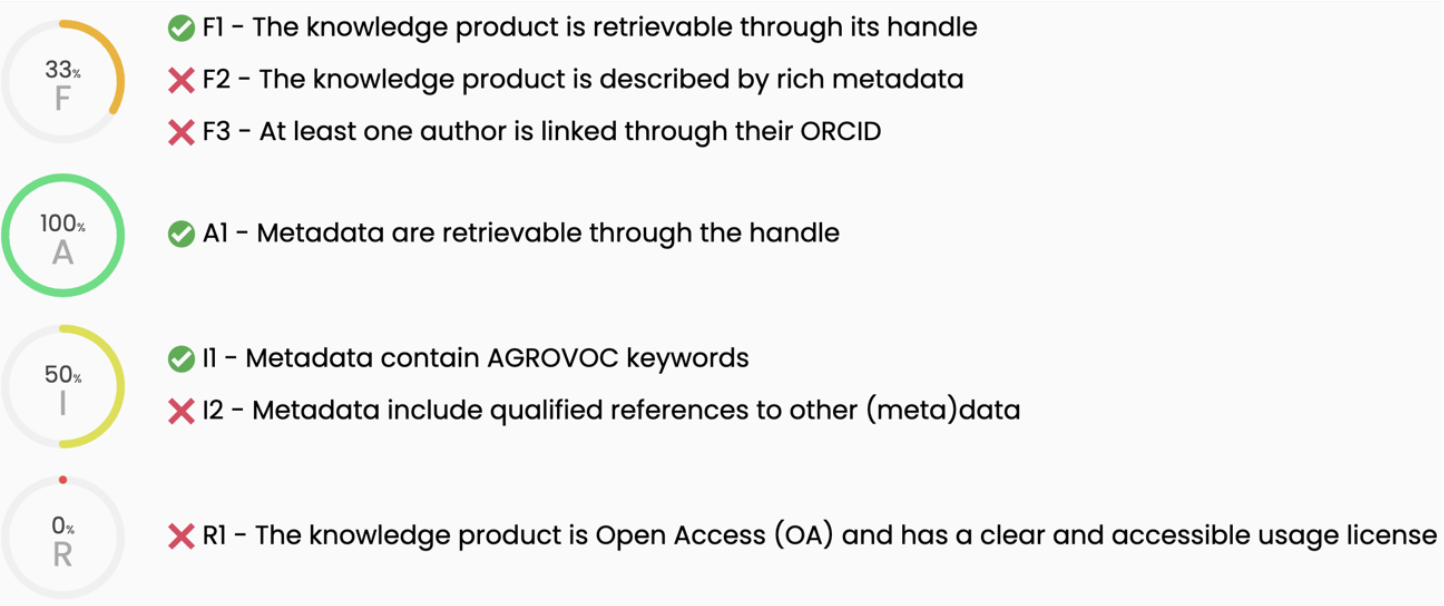}
    \caption{An example of automated FAIR scoring and display in agricultural science at CGIAR.}
    \label{fig:fairscoring}
\end{figure}

Researchers attempt to create automated FAIR scoring algorithms in response to the increasing volume of scientific data. For example, FAIR EVA~\cite{aguilar2022fair} is developed to evaluate, validate, and advise on FAIR compliance. \autoref{fig:fairscoring} provides a concrete example from global agricultural science on adopting automated FAIR scoring in the reporting tool of CGIAR, an international partnership of organizations dedicated to sustainable food production. While the development of automated FAIR scoring systems is at an early stage, HCI might profit from early adoption. In particular, FAIR scoring algorithms might, that way, be early to consider the broad spectrum of research data used in a community like ours. In return, FAIR score developers will profit from the involvement of HCI researchers and practitioners by learning about the user experience of scientists confronted with these novel scores.

\section{Ways to Move Forward}

Given all the activities, initiatives, and projects sketched above, the question is how to move forward. This is to create awareness and readiness supporting proper RDM in normal scientific processes within HCI. We believe it needs a combined bottom-up and top-down approach as well as creating bonds across national boundaries to connect the various existing infrastructures properly.

Bottom-up means that comprehensive education, advice, stewardship, etc., supporting the daily work of scientists based on incentives for them, while top-down needs, for example, requirements for funding proposals to include RDM-related work packages as well as improved/extended publication processes for scientific outlets to include research data and related material (e.g., entirely fair digital objects like an RDMC) for a proper publication.

\subsection{Incentives} 
RDM must be made attractive in its own right so that scientists start picking it up voluntarily. Next to such measures discussed below, certain collateral activities like brief and comprehensive (online) lessons to support their daily work need to be established. There are a few small courses available\footnote{\url{https://learning.theodi.org/courses}; last access: 2023-07-12}\footnote{\url{https://eo4society.esa.int/projects/mooc-eo-open-data-science/}; last access: 2023-07-12}, with more coming soon. Here, the community is asked to provide HCI-specific input to tailor generic RDM courses to make HCI RDM a breeze. 

The benefits of providing the related research data packaged within a publication will certainly be better visibility. Early experience shows that research results coming with the research data are (much) more likely to be cited~\cite{piwowar2013data}. In addition, once a proper citation scheme for data and software is globally established, contributions to the research results of others using software components or reusing the research data will add to the authors' visibility, reputation, and citation count.

The result will also be higher productivity. This has many reasons. One is that the review processes for the research data (see also below) will ensure that it is easier to decide which work (i.e., data) is worthwhile a try and can be integrated into one's own research. Furthermore, the related software components to purge, analyze, and present the data can be reused more easily. Thus, reusable research data is a springboard for starting new research ideas. This can also be improved through funding schemes that allow getting the support of data stewards or similar consultancy schemes improving the RDM competencies.  

\subsection{Mandates} 
There are two areas where one can and, in some instances, has to expect already now certain requirements that RDM is performed as part of the normal scientific processes. These are research funding and quality assurance processes (reviews) for scientific outlets like journals and conferences.

Funding regulations are currently (re)designed to include RDM as compulsory components in many organizations. As an example, the German basic research funding agency issued a statement defining that any project proposal needs a proper definition of work packages regarding the project/topic-specific RDM in spring 2022~\cite{dfg1}. While it was already a strong recommendation and, in some cases, an obligatory element in some types of proposals, it is now necessary for all proposals handed into the DFG. Similar situations can be found in the UK, The Netherlands, France, the USA, etc. Thus, one has to expect more requirements to come, and they will get more specific over time. Here, the HCI community needs to help define the specific requirements of its field to meet the requirements of better HCI research.

The other aspect—reviewing publications, including the research data—will also require special attention (e.g., see \cite{ArtifactTransp2020}). The definition of quality criteria for the various types of HCI research data types mentioned above will be a community effort and will evolve as review processes currently do anyway. Thus, the inclusion of research data and related context or artifacts will need specific attention here. 
Additionally, it is noteworthy that the additional review load needs to be addressed. Generally, it is felt that the review load is already high, and additional work needed to review the accompanying research data will certainly add to that. The amelioration of this aspect needs HCI-specific effort to remove this as an obstacle to requiring RDM for publications. A few measures can be applied here, like standard ways to present data and metadata, the aforementioned quality criteria as well as some automation in the assessment processes. This, of course, depends to a great extent on the research data type in question: tables of measured data can be processed by statistics tools much more easily compared to the qualitative data, as sketched above. 

Furthermore, certain aspects should be agreed upon by the HCI community to remove the negative incentives for sharing research data:
\begin{itemize}
    \item The HCI community should agree that sharing a certain research data set (combined with software and other context information in an RDMC, for example) does not preempt a publication of a related paper. The research data can stand on its own and has its own persistent identifier, like a DOI. 
    \item Make it possible to hold data in escrow so that authors do not need to worry about others publishing their data before their work based on that data is published. Again a proper scheme of referencing research data will be instrumental for this kind of feature
    \item There should be enough space to store research data containers and computing power to resurrect them for replication/validation studies. This is because not all research data might find a home in the existing/emerging infrastructures.
    \item Replication studies should not be avoided but encouraged
\end{itemize}

Various infrastructure and processes are available, e.g., \url{https://zenodo.org} or \url{https://arxiv.org}.
Below, we summarize and briefly discuss our questions and proposed ways to proceed.







\subsection{Discussion}
Given all these nice-to-have properties, the question is how to achieve these and how to implement these in a way to have a smooth and swift transition to a better inclusion of RDM in HCI research.
This should be done also through careful consideration of the costs and benefits. Since there is some funding available in the system, it will be a great chance to use the existing expertise to create tool support to help with automating at least some of the boring aspects of the work — especially reducing the costs mentioned.

Furthermore, it is essential that these efforts be led by the scientific community, i.e., HCI RDM profile is defined by the  HCI community. This is important that the needs of the HCI community are addressed and not part of a profit-oriented stakeholder or some other body outside the science community.

Here come the scientific societies like ACM, IEEE CS, IFIP, and all the national CS societies like the BCS in the UK, GI in Germany, and their related subgroups of HCI specialists to arrange for discussions and processes leading to agreement on, e.g., metadata types, quality criteria and interoperability of infrastructures. Flagship conference series like CHI in HCI play a leading role in serving as a source of related knowledge and competencies as well as a bed to implement ideas and concepts.

Of course, it is a good idea to start with existing activities like the badge system and the FAIR-scoring scheme and put them into their respective places in the greater picture sketched above. The results of activities from the RepliCHI efforts or transparent statistics are then important ingredients to consider, especially in the HCI context. To complete the picture, the other aspects to support reuse and long-term archival to support the FAIR principles for HCI fully will probably overcome the effects shown in~\cite{ArtifactTransp2020}. 





\section{Conclusion}
We raised the issue of RDM in the research area of HCI to improve the reusability and quality of publications. The main purpose is to increase the sustainability and credibility of research results and make research more cost-effective and productive by sharing research data. We sketched several aspects characterizing research data types in HCI and existing and emerging infrastructures for storing them.

Overall, the need to establish RDM in HCI is becoming more important. Thus, it needs to become a topic for discussion to find agreement regarding quality criteria, processes, and structures within the HCI discipline, which also crosses borders of scientific societies (ACM, IEEE, etc.), domains of funding agencies, and nationally bound efforts.

A number of issues and questions have been collected and efforts will be made to pick these up to address the topics of quality and process in the framework of the FAIR principles for HCI.
\begin{acks}
    This work was supported by funding from the National Science Foundation (NSF) under grant number IIS-2107111. 
\end{acks}

\bibliographystyle{ACM-Reference-Format}
\bibliography{bibliography.bib}


\begin{thebibliography}{37}


\ifx \showCODEN    \undefined \def \showCODEN     #1{\unskip}     \fi
\ifx \showISBNx    \undefined \def \showISBNx     #1{\unskip}     \fi
\ifx \showISBNxiii \undefined \def \showISBNxiii  #1{\unskip}     \fi
\ifx \showISSN     \undefined \def \showISSN      #1{\unskip}     \fi
\ifx \showLCCN     \undefined \def \showLCCN      #1{\unskip}     \fi
\ifx \shownote     \undefined \def \shownote      #1{#1}          \fi
\ifx \showarticletitle \undefined \def \showarticletitle #1{#1}   \fi
\ifx \showURL      \undefined \def \showURL       {\relax}        \fi
\providecommand\bibfield[2]{#2}
\providecommand\bibinfo[2]{#2}
\providecommand\natexlab[1]{#1}
\providecommand\showeprint[2][]{arXiv:#2}

\bibitem[Accessibility(2020)]%
        {w4a}
\bibfield{author}{\bibinfo{person}{W4A:~Web Accessibility}.}
  \bibinfo{year}{2020}\natexlab{}.
\newblock \bibinfo{title}{{W4A: Web Accessibility}}.
\newblock \bibinfo{howpublished}{\url{https://dl.acm.org/conference/w4a}}.
\newblock
\newblock
\shownote{[Online; accessed 12-JUNE-2022]}.


\bibitem[Aguilar(2022)]%
        {aguilar2022fair}
\bibfield{author}{\bibinfo{person}{Fernando Aguilar}.}
  \bibinfo{year}{2022}\natexlab{}.
\newblock \bibinfo{title}{FAIR EVA-EGI 2022 Demo}.
\newblock


\bibitem[Chevalier et~al\mbox{.}(2023)]%
        {transparentstatistics}
\bibfield{author}{\bibinfo{person}{Fanny Chevalier}, \bibinfo{person}{Lewis
  Chuang}, \bibinfo{person}{Pierre Dragicevic}, \bibinfo{person}{Shion Guha},
  \bibinfo{person}{Steve Haroz}, \bibinfo{person}{Matthew Kay}, {and}
  \bibinfo{person}{Chat Wacharamanotham}.} \bibinfo{year}{2023}\natexlab{}.
\newblock \bibinfo{title}{Transparent Statistics in Human–Computer
  Interaction}.
\newblock
\urldef\tempurl%
\url{https://transparentstatistics.org/}
\showURL{%
Retrieved January 19, 2023 from \tempurl}


\bibitem[Chuang and Pfeil(2018)]%
        {10.1145/3170427.3185377}
\bibfield{author}{\bibinfo{person}{Lewis~L. Chuang} {and}
  \bibinfo{person}{Ulrike Pfeil}.} \bibinfo{year}{2018}\natexlab{}.
\newblock \showarticletitle{Transparency and Openness Promotion Guidelines for
  HCI}. In \bibinfo{booktitle}{\emph{Extended Abstracts of the 2018 CHI
  Conference on Human Factors in Computing Systems}} (Montreal QC, Canada)
  \emph{(\bibinfo{series}{CHI EA '18})}. \bibinfo{publisher}{Association for
  Computing Machinery}, \bibinfo{address}{New York, NY, USA},
  \bibinfo{pages}{1–4}.
\newblock
\showISBNx{9781450356213}
\href{https://doi.org/10.1145/3170427.3185377}{doi:\nolinkurl{10.1145/3170427.3185377}}


\bibitem[Colley et~al\mbox{.}(2022a)]%
        {10.1145/3491102.3517571}
\bibfield{author}{\bibinfo{person}{Mark Colley}, \bibinfo{person}{Elvedin
  Bajrovic}, {and} \bibinfo{person}{Enrico Rukzio}.}
  \bibinfo{year}{2022}\natexlab{a}.
\newblock \showarticletitle{Effects of Pedestrian Behavior, Time Pressure, and
  Repeated Exposure on Crossing Decisions in Front of Automated Vehicles
  Equipped with External Communication}. In
  \bibinfo{booktitle}{\emph{Proceedings of the 2022 CHI Conference on Human
  Factors in Computing Systems}} (New Orleans, LA, USA)
  \emph{(\bibinfo{series}{CHI '22})}. \bibinfo{publisher}{Association for
  Computing Machinery}, \bibinfo{address}{New York, NY, USA}, Article
  \bibinfo{articleno}{367}, \bibinfo{numpages}{11}~pages.
\newblock
\showISBNx{9781450391573}
\href{https://doi.org/10.1145/3491102.3517571}{doi:\nolinkurl{10.1145/3491102.3517571}}


\bibitem[Colley et~al\mbox{.}(2022b)]%
        {colley2022requirements}
\bibfield{author}{\bibinfo{person}{Mark Colley}, \bibinfo{person}{Stefanos
  Mytilineos}, \bibinfo{person}{Marcel Walch}, \bibinfo{person}{Jan
  Gugenheimer}, {and} \bibinfo{person}{Enrico Rukzio}.}
  \bibinfo{year}{2022}\natexlab{b}.
\newblock \showarticletitle{Requirements for the interaction with highly
  automated construction site delivery trucks}.
\newblock \bibinfo{journal}{\emph{Frontiers in Human Dynamics}}
  \bibinfo{volume}{4} (\bibinfo{year}{2022}), \bibinfo{numpages}{12}~pages.
\newblock


\bibitem[Committee(2023)]%
        {fairdo}
\bibfield{author}{\bibinfo{person}{FAIRDO Forum~Steering Committee}.}
  \bibinfo{year}{2023}\natexlab{}.
\newblock \bibinfo{title}{Fair Digital Objet Forum}.
\newblock
\urldef\tempurl%
\url{https://fairdo.org/}
\showURL{%
Retrieved January 19, 2023 from \tempurl}


\bibitem[DataScience(2023)]%
        {NFDI4ds}
\bibfield{author}{\bibinfo{person}{NFDI~4 DataScience}.}
  \bibinfo{year}{2023}\natexlab{}.
\newblock \bibinfo{title}{NFDI 4 DataScience}.
\newblock
\urldef\tempurl%
\url{https://www.nfdi4datascience.de/}
\showURL{%
Retrieved January 19, 2023 from \tempurl}


\bibitem[Davis and Connolly(2007)]%
        {davis2007institutional}
\bibfield{author}{\bibinfo{person}{Philip~M Davis} {and}
  \bibinfo{person}{Matthew~JL Connolly}.} \bibinfo{year}{2007}\natexlab{}.
\newblock \showarticletitle{Institutional repositories: evaluating the reasons
  for non-use of Cornell University's installation of DSpace}.
\newblock \bibinfo{journal}{\emph{D-Lib Magazine}}  \bibinfo{volume}{13}
  (\bibinfo{year}{2007}), \bibinfo{numpages}{17}~pages.
\newblock


\bibitem[DFG(2022a)]%
        {dfg2}
\bibfield{author}{\bibinfo{person}{DFG}.} \bibinfo{year}{2022}\natexlab{a}.
\newblock \bibinfo{booktitle}{\emph{Nationale Forschungsdateninfrastruktur}}.
\newblock DFG.
\newblock
\urldef\tempurl%
\url{https://www.dfg.de/nfdi}
\showURL{%
Retrieved January 19, 2023 from \tempurl}


\bibitem[DFG(2022b)]%
        {dfg1}
\bibfield{author}{\bibinfo{person}{DFG}.} \bibinfo{year}{2022}\natexlab{b}.
\newblock \bibinfo{booktitle}{\emph{Specification of Requirements Relating to
  the Handling of Research Data in Funding Proposals}}.
\newblock DFG.
\newblock
\urldef\tempurl%
\url{https://www.dfg.de/en/research_funding/announcements_proposals/2022/info_wissenschaft_22_25/index.html}
\showURL{%
Retrieved January 19, 2023 from \tempurl}


\bibitem[Fair(2023)]%
        {gofair}
\bibfield{author}{\bibinfo{person}{GO Fair}.} \bibinfo{year}{2023}\natexlab{}.
\newblock \bibinfo{title}{FAIR Principles}.
\newblock
\urldef\tempurl%
\url{https://www.go-fair.org/fair-principles/}
\showURL{%
Retrieved January 19, 2023 from \tempurl}


\bibitem[Feger et~al\mbox{.}(2019)]%
        {10.1145/3290607.3312905}
\bibfield{author}{\bibinfo{person}{Sebastian~S. Feger},
  \bibinfo{person}{S\"{u}nje Dallmeier-Tiessen}, \bibinfo{person}{Pawe\l{}~W.
  Wo\'{z}niak}, {and} \bibinfo{person}{Albrecht Schmidt}.}
  \bibinfo{year}{2019}\natexlab{}.
\newblock \showarticletitle{The Role of HCI in Reproducible Science:
  Understanding, Supporting and Motivating Core Practices}. In
  \bibinfo{booktitle}{\emph{Extended Abstracts of the 2019 CHI Conference on
  Human Factors in Computing Systems}} (Glasgow, Scotland Uk)
  \emph{(\bibinfo{series}{CHI EA '19})}. \bibinfo{publisher}{Association for
  Computing Machinery}, \bibinfo{address}{New York, NY, USA},
  \bibinfo{pages}{1–6}.
\newblock
\showISBNx{9781450359719}
\href{https://doi.org/10.1145/3290607.3312905}{doi:\nolinkurl{10.1145/3290607.3312905}}


\bibitem[Feger et~al\mbox{.}(2021)]%
        {tailored_badges}
\bibfield{author}{\bibinfo{person}{Sebastian~Stefan Feger},
  \bibinfo{person}{Pawe\l{}~W. Wo\'{z}niak}, \bibinfo{person}{Jasmin Niess},
  {and} \bibinfo{person}{Albrecht Schmidt}.} \bibinfo{year}{2021}\natexlab{}.
\newblock \showarticletitle{Tailored Science Badges: Enabling New Forms of
  Research Interaction}. In \bibinfo{booktitle}{\emph{Designing Interactive
  Systems Conference 2021}} (Virtual Event, USA) \emph{(\bibinfo{series}{DIS
  '21})}. \bibinfo{publisher}{Association for Computing Machinery},
  \bibinfo{address}{New York, NY, USA}, \bibinfo{pages}{576–588}.
\newblock
\showISBNx{9781450384766}
\href{https://doi.org/10.1145/3461778.3462067}{doi:\nolinkurl{10.1145/3461778.3462067}}


\bibitem[Foundation(2008)]%
        {NSFROBOTS}
\bibfield{author}{\bibinfo{person}{National~Science Foundation}.}
  \bibinfo{year}{2008}\natexlab{}.
\newblock \bibinfo{booktitle}{\emph{Cyberinfrastructure for Human-Robot
  Interaction Research}}.
\newblock NFS.
\newblock
\urldef\tempurl%
\url{https://www.nsf.gov/awardsearch/showAward?AWD_ID=0742350}
\showURL{%
Retrieved January 19, 2023 from \tempurl}


\bibitem[Foundation(2023)]%
        {NSFCyber}
\bibfield{author}{\bibinfo{person}{National~Science Foundation}.}
  \bibinfo{year}{2023}\natexlab{}.
\newblock \bibinfo{title}{Cyberinfrastructure for Sustained Scientific
  Innovation (CSSI)Cyberinfrastructure for Sustained Scientific Innovation
  (CSSI)}.
\newblock
\urldef\tempurl%
\url{https://beta.nsf.gov/funding/opportunities/cyberinfrastructure-sustained-scientifichttps://beta.nsf.gov/funding/opportunities/cyberinfrastructure-sustained-scientific}
\showURL{%
Retrieved January 19, 2023 from \tempurl}


\bibitem[generic et~al\mbox{.}(2023)]%
        {qemu}
\bibfield{author}{\bibinfo{person}{QEMU~A generic}, \bibinfo{person}{open
  source~machine emulator}, {and} \bibinfo{person}{virtualizer}.}
  \bibinfo{year}{2023}\natexlab{}.
\newblock \bibinfo{title}{{QEMU A generic and open source machine emulator and
  virtualizer}}.
\newblock
\urldef\tempurl%
\url{https://www.qemu.org}
\showURL{%
Retrieved January 19, 2023 from \tempurl}


\bibitem[Goedicke and Lucke(2022)]%
        {mci/Goedicke2022}
\bibfield{author}{\bibinfo{person}{Michael Goedicke} {and}
  \bibinfo{person}{Ulrike Lucke}.} \bibinfo{year}{2022}\natexlab{}.
\newblock \showarticletitle{Research Data Management in Computer Science -
  NFDIxCS Approach}. In \bibinfo{booktitle}{\emph{INFORMATIK 2022}},
  \bibfield{editor}{\bibinfo{person}{Daniel Demmler}, \bibinfo{person}{Daniel
  Krupka}, {and} \bibinfo{person}{Hannes Federrath}} (Eds.).
  \bibinfo{publisher}{Gesellschaft für Informatik}, \bibinfo{address}{Bonn,
  Germany}, \bibinfo{pages}{1317--1328}.
\newblock
\href{https://doi.org/10.18420/inf2022_112}{doi:\nolinkurl{10.18420/inf2022_112}}


\bibitem[Interfaces and Applications(2020)]%
        {autoui}
\bibfield{author}{\bibinfo{person}{AutomotiveUI: Automotive~User Interfaces}
  {and} \bibinfo{person}{Interactive~Vehicular Applications}.}
  \bibinfo{year}{2020}\natexlab{}.
\newblock \bibinfo{title}{{AutomotiveUI: Automotive User Interfaces and
  Interactive Vehicular Applications}}.
\newblock
  \bibinfo{howpublished}{\url{https://dl.acm.org/conference/automotiveui}}.
\newblock
\newblock
\shownote{[Online; accessed 12-JUNE-2022]}.


\bibitem[Ive et~al\mbox{.}(2015)]%
        {10.1145/2809730.2809752}
\bibfield{author}{\bibinfo{person}{Hillary~Page Ive}, \bibinfo{person}{David
  Sirkin}, \bibinfo{person}{Dave Miller}, \bibinfo{person}{Jamy Li}, {and}
  \bibinfo{person}{Wendy Ju}.} \bibinfo{year}{2015}\natexlab{}.
\newblock \showarticletitle{"Don't Make Me Turn This Seat around!": Driver and
  Passenger Activities and Positions in Autonomous Cars}. In
  \bibinfo{booktitle}{\emph{Adjunct Proceedings of the 7th International
  Conference on Automotive User Interfaces and Interactive Vehicular
  Applications}} (Nottingham, United Kingdom)
  \emph{(\bibinfo{series}{AutomotiveUI '15})}. \bibinfo{publisher}{Association
  for Computing Machinery}, \bibinfo{address}{New York, NY, USA},
  \bibinfo{pages}{50–55}.
\newblock
\showISBNx{9781450338585}
\href{https://doi.org/10.1145/2809730.2809752}{doi:\nolinkurl{10.1145/2809730.2809752}}


\bibitem[Kidwell et~al\mbox{.}(2016)]%
        {kidwell2016badges}
\bibfield{author}{\bibinfo{person}{Mallory~C Kidwell},
  \bibinfo{person}{Ljiljana~B Lazarevi{\'c}}, \bibinfo{person}{Erica Baranski},
  \bibinfo{person}{Tom~E Hardwicke}, \bibinfo{person}{Sarah Piechowski},
  \bibinfo{person}{Lina-Sophia Falkenberg}, \bibinfo{person}{Curtis Kennett},
  \bibinfo{person}{Agnieszka Slowik}, \bibinfo{person}{Carina Sonnleitner},
  \bibinfo{person}{Chelsey Hess-Holden}, {et~al\mbox{.}}}
  \bibinfo{year}{2016}\natexlab{}.
\newblock \showarticletitle{Badges to acknowledge open practices: A simple,
  low-cost, effective method for increasing transparency}.
\newblock \bibinfo{journal}{\emph{PLoS biology}} \bibinfo{volume}{14},
  \bibinfo{number}{5} (\bibinfo{year}{2016}), \bibinfo{pages}{e1002456}.
\newblock


\bibitem[LeBeau(2019)]%
        {brandon2019package}
\bibfield{author}{\bibinfo{person}{Brandon LeBeau}.}
  \bibinfo{year}{2019}\natexlab{}.
\newblock \bibinfo{title}{Package ‘pdfsearch’}.
\newblock
\urldef\tempurl%
\url{https://cran.r-project.org/web/packages/pdfsearch/pdfsearch.pdf}
\showURL{%
\tempurl}


\bibitem[Libraries(2023)]%
        {dspace}
\bibfield{author}{\bibinfo{person}{MIT Libraries}.}
  \bibinfo{year}{2023}\natexlab{}.
\newblock \bibinfo{title}{DSpace at MIT}.
\newblock
\urldef\tempurl%
\url{http://dspace.mit.edu/}
\showURL{%
Retrieved January 19, 2023 from \tempurl}


\bibitem[Library(2023)]%
        {eommons}
\bibfield{author}{\bibinfo{person}{Cornell~University Library}.}
  \bibinfo{year}{2023}\natexlab{}.
\newblock \bibinfo{title}{eCommons Open scholarship at Cornell}.
\newblock
\urldef\tempurl%
\url{https://ecommons.cornell.edu}
\showURL{%
Retrieved January 19, 2023 from \tempurl}


\bibitem[Mobile and Multimedia(2020)]%
        {mum}
\bibfield{author}{\bibinfo{person}{MUM: Mobile} {and}
  \bibinfo{person}{Ubiquitous Multimedia}.} \bibinfo{year}{2020}\natexlab{}.
\newblock \bibinfo{title}{{MUM: Mobile and Ubiquitous Multimedia}}.
\newblock \bibinfo{howpublished}{\url{https://dl.acm.org/conference/mum}}.
\newblock
\newblock
\shownote{[Online; accessed 12-JUNE-2022]}.


\bibitem[nfdixcs(2023)]%
        {nfdixcs}
\bibfield{author}{\bibinfo{person}{nfdixcs}.} \bibinfo{year}{2023}\natexlab{}.
\newblock \bibinfo{title}{National Research Data Infrastructure for and with
  Computer Science (NFDIxCS)}.
\newblock
\urldef\tempurl%
\url{https://www.nfdixcs.org}
\showURL{%
Retrieved January 19, 2023 from \tempurl}


\bibitem[on~Accessible~Computing(2020)]%
        {taccess}
\bibfield{author}{\bibinfo{person}{ACM~Transactions on Accessible~Computing}.}
  \bibinfo{year}{2020}\natexlab{}.
\newblock \bibinfo{title}{{ACM Transactions on Accessible Computing}}.
\newblock \bibinfo{howpublished}{\url{https://dl.acm.org/journal/taccess}}.
\newblock
\newblock
\shownote{[Online; accessed 12-JUNE-2020]}.


\bibitem[on~Human Factors~in Computing~Systems(2020)]%
        {chi}
\bibfield{author}{\bibinfo{person}{CHI:~Conference on~Human Factors~in
  Computing~Systems}.} \bibinfo{year}{2020}\natexlab{}.
\newblock \bibinfo{title}{{CHI: Conference on Human Factors in Computing
  Systems}}.
\newblock \bibinfo{howpublished}{\url{https://dl.acm.org/conference/chi}}.
\newblock
\newblock
\shownote{[Online; accessed 12-JUNE-2022]}.


\bibitem[Parliament and the Council of~the European~Union(2016)]%
        {eu2018gdpr}
\bibfield{author}{\bibinfo{person}{The~European Parliament} {and}
  \bibinfo{person}{the Council of~the European~Union}.}
  \bibinfo{year}{2016}\natexlab{}.
\newblock \bibinfo{title}{General Data Protection Regulation}.
\newblock
\urldef\tempurl%
\url{https://eur-lex.europa.eu/eli/reg/2016/679/oj}
\showURL{%
Retrieved May 13 2021 from \tempurl}


\bibitem[Piwowar and Vision(2013)]%
        {piwowar2013data}
\bibfield{author}{\bibinfo{person}{Heather~A Piwowar} {and}
  \bibinfo{person}{Todd~J Vision}.} \bibinfo{year}{2013}\natexlab{}.
\newblock \showarticletitle{Data reuse and the open data citation advantage}.
\newblock \bibinfo{journal}{\emph{PeerJ}}  \bibinfo{volume}{1}
  (\bibinfo{year}{2013}), \bibinfo{pages}{e175}.
\newblock


\bibitem[Reid(2008)]%
        {reid2008investigation}
\bibfield{author}{\bibinfo{person}{Stephanie Reid}.}
  \bibinfo{year}{2008}\natexlab{}.
\newblock \emph{\bibinfo{title}{An investigation into the motivating factors
  behind the use or non use of institutional repositories by selected
  university academics}}.
\newblock \bibinfo{thesistype}{Ph.\,D. Dissertation}.
  \bibinfo{school}{ResearchArchive}.
\newblock


\bibitem[Software and Technology(2020)]%
        {uist}
\bibfield{author}{\bibinfo{person}{UIST: User~Interface Software} {and}
  \bibinfo{person}{Technology}.} \bibinfo{year}{2020}\natexlab{}.
\newblock \bibinfo{title}{{UIST: User Interface Software and Technology}}.
\newblock \bibinfo{howpublished}{\url{https://dl.acm.org/conference/uist}}.
\newblock
\newblock
\shownote{[Online; accessed 12-JUNE-2020]}.


\bibitem[Stanford-Libraries(2023)]%
        {stanford}
\bibfield{author}{\bibinfo{person}{Stanford-Libraries}.}
  \bibinfo{year}{2023}\natexlab{}.
\newblock \bibinfo{title}{Stanford Digital Repository}.
\newblock
\urldef\tempurl%
\url{https://sdr.stanford.edu/}
\showURL{%
Retrieved January 19, 2023 from \tempurl}


\bibitem[surf.nl(2023)]%
        {surf}
\bibfield{author}{\bibinfo{person}{surf.nl}.} \bibinfo{year}{2023}\natexlab{}.
\newblock \bibinfo{title}{National Coordination Point Research Data
  Management}.
\newblock
\urldef\tempurl%
\url{https://www.surf.nl/en/national-coordination-point-research-data-management}
\showURL{%
Retrieved January 19, 2023 from \tempurl}


\bibitem[Wacharamanotham et~al\mbox{.}(2020)]%
        {ArtifactTransp2020}
\bibfield{author}{\bibinfo{person}{Chat Wacharamanotham},
  \bibinfo{person}{Lukas Eisenring}, \bibinfo{person}{Steve Haroz}, {and}
  \bibinfo{person}{Florian Echtler}.} \bibinfo{year}{2020}\natexlab{}.
\newblock \showarticletitle{Transparency of CHI Research Artifacts: Results of
  a Self-Reported Survey}. In \bibinfo{booktitle}{\emph{Proceedings of the 2020
  CHI Conference on Human Factors in Computing Systems}} (Honolulu, HI, USA)
  \emph{(\bibinfo{series}{CHI '20})}. \bibinfo{publisher}{Association for
  Computing Machinery}, \bibinfo{address}{New York, NY, USA},
  \bibinfo{pages}{1–14}.
\newblock
\showISBNx{9781450367080}
\href{https://doi.org/10.1145/3313831.3376448}{doi:\nolinkurl{10.1145/3313831.3376448}}


\bibitem[Wilson et~al\mbox{.}(2011)]%
        {10.1145/1979742.1979491}
\bibfield{author}{\bibinfo{person}{Max~L. Wilson}, \bibinfo{person}{Wendy
  Mackay}, \bibinfo{person}{Ed Chi}, \bibinfo{person}{Michael Bernstein},
  \bibinfo{person}{Dan Russell}, {and} \bibinfo{person}{Harold Thimbleby}.}
  \bibinfo{year}{2011}\natexlab{}.
\newblock \showarticletitle{RepliCHI - CHI Should Be Replicating and Validating
  Results More: Discuss}. In \bibinfo{booktitle}{\emph{CHI '11 Extended
  Abstracts on Human Factors in Computing Systems}} (Vancouver, BC, Canada)
  \emph{(\bibinfo{series}{CHI EA '11})}. \bibinfo{publisher}{Association for
  Computing Machinery}, \bibinfo{address}{New York, NY, USA},
  \bibinfo{pages}{463–466}.
\newblock
\showISBNx{9781450302685}
\href{https://doi.org/10.1145/1979742.1979491}{doi:\nolinkurl{10.1145/1979742.1979491}}


\bibitem[Zhou et~al\mbox{.}(2022)]%
        {10.1145/3534587}
\bibfield{author}{\bibinfo{person}{Hao Zhou}, \bibinfo{person}{Taiting Lu},
  \bibinfo{person}{Yilin Liu}, \bibinfo{person}{Shijia Zhang}, {and}
  \bibinfo{person}{Mahanth Gowda}.} \bibinfo{year}{2022}\natexlab{}.
\newblock \showarticletitle{Learning on the Rings: Self-Supervised 3D Finger
  Motion Tracking Using Wearable Sensors}.
\newblock \bibinfo{journal}{\emph{Proc. ACM Interact. Mob. Wearable Ubiquitous
  Technol.}} \bibinfo{volume}{6}, \bibinfo{number}{2}, Article
  \bibinfo{articleno}{90} (\bibinfo{date}{jul} \bibinfo{year}{2022}),
  \bibinfo{numpages}{31}~pages.
\newblock
\href{https://doi.org/10.1145/3534587}{doi:\nolinkurl{10.1145/3534587}}


\end{thebibliography}

\end{document}